\documentstyle[12pt,twoside]{article}
\begin{document}
\begin{center}

{\Large\bf 
DENSITY PERTURBATIONS IN AN UNIVERSE\\[5PT]
DOMINATED BY THE CHAPLYGIN GAS\\[5PT]}
\medskip
 
{\bf
J. C. Fabris\footnote{e-mail: fabris@cce.ufes.br},
S.V.B. Gon\c{c}alves\footnote{e-mail: sergio@cce.ufes.br} and P.E. de Souza\footnote{e-mail: patricia.ilus@bol.com.br}}  \medskip

Departamento de F\'{\i}sica, Universidade Federal do Esp\'{\i}rito Santo, 
CEP29060-900, Vit\'oria, Esp\'{\i}rito Santo, Brazil \medskip

\end{center}
 
\begin{abstract}
\vspace{0.7cm}
We study the fate of density perturbations in an Universe dominate
by the Chaplygin gas, which exhibit negative pressure. We show that it is possible to obtain the
value for the
density contrast observed in large scale structure of the Universe
by fixing a free parameter in the equation of state of this gas.
The negative character of pressure must be significant only very
recently.
\newline
PACS number(s): 98.80.Bp, 98.65.Dx
\end{abstract}

The nature of "missing mass" that must dominate the Universe today is one
of the most intriguing problem in modern physics. This problem
appears at various levels, beginning with the rotation curve of spiral
galaxies \cite{florido}, and extending to the total mass of the Universe as
it is deduced through the first acoustic peak of the spectrum of
the anisotropy of cosmic background radiation \cite{frampton}.
A fraction of this missing mass has been called dark energy since
it does cluster, remaining a smooth component in the Universe.
This problem
has acquired a most dramatic face through the results coming from
the observation of supernova type Ia, which indicate that the Universe
today is in an accelerated expansion \cite{riess,perlmu1}. Hence, the inflationary paradigma,
first conceived for the primordial Universe, can be
applicable to our actual Universe.
\par
In this way, the cosmological constant problem was revived: it is possible
that a cosmological constant may be the dominant component of matter in
the Universe. However, other types of matter with negative pressure have
been considered as candidate of the so-called dark energy of the Universe:
fluids of topological defects (domain walls, strings), quintessence
(a self-interacting scalar field minimally coupled to gravity), etc.
Recently, a different matter component with negative pressure has
been presented as possible candidate for this dark energy: the
Chaplygin gas \cite{pasquier}.
\par
The Chaplygin gas has an equation of state of the type
\begin{equation}
\label{pressao}
p = - \frac{A}{\rho} \quad ,
\end{equation}
where $A$ is a constant. This exotic fluid has an interesting origin.
Taking the Nambu-Goto action of string theory, the Chaplygin
fluid appears after considering $d$-branes in a $d + 2$ dimensional
space-time, in the light cone gauge \cite{hoppe}.
In this case, the Nambu action reduces to \cite{ogawa}
\begin{equation}
\label{acao}
S = \int d^{d+1}r\biggr[\theta\dot\rho - \biggr(\rho(\nabla\theta)^2 +
\frac{1}{2}\frac{A}{\rho}\biggl)\biggl] \quad .
\end{equation}
A very curious aspect of this action is that the variables $\theta$
and $\rho$ obey the newtonian hydrodynamic equations for an
irrotational fluid with
the pressure given by (\ref{pressao}). Moreover, the Chaplygin gas
admit a supersymmetric extension.
\par
In \cite{pasquier}, the consequences for the evolution of the Universe
of such exotic fluid have been investigate.
They found that, in the relativistic context, the density is connected to
the scale factor as
\begin{equation}
\rho = \sqrt{A + \frac{B}{a^6}} \quad .
\end{equation}
Hence, initially the Universe behaves as it was dominated by a dust
fluid, and than the density becomes asymptotically constant, revealing
that the cosmological constant becomes the dominant component of the Universe. There is an intermediate phase, which can be described by
a cosmological constant mixed with a stiff matter fluid.
In this sense, this exotic fluid may be a candidate for describing
the Universe as the supernova observational results indicate it must be.
\par
In the presente note, we study another aspect of this fluid: the
evolution of density perturbations. A relativistic study of this
problem leads to some technical difficulties. Considering that
we can approximate the Universe filled by such a fluid with the
three phases described before, the behaviour of density perturbation
in the initial and final phases are already known ($\delta_{dust} \propto
t^{2/3}$ and $\delta_{cc} = 0$ respectively). Hence in order to obtain non-trivial
effects we must consider the intermediate phase. However, such a phase
admits solution for the background, but no analytical solution
for the perturbed equations.
\par
Since our intention is to have explicity expression for density
perturbations, in order at least to evaluate what happens for the
evolution of structure in an Universe dominate by the Chaplygin gas,
we will
exploit the fact that the action (\ref{acao}) leads to the
hydrodynamic newtonian
equations.
Coupling to (\ref{acao}) a term representing the gravitational potential $\phi$
we have the following equations \cite{weinberg}:
\begin{eqnarray}
\label{be1}
\frac{\partial\rho}{\partial t} + \nabla.(\rho\vec v) &=& 0 \quad ,\\
\label{be2}
\dot{\vec v} + \vec v.\nabla\vec v &=& - \frac{\nabla p}{\rho} +
\nabla\phi \quad , \\
\label{be3}
\nabla^2\phi &=& - 4\pi G\rho \quad .
\end{eqnarray}
If these equations describe an expanding homogeneous and isotropic Universe
their solution is
\begin{equation}
\rho = \frac{\rho_0}{a^3} \quad , \quad \vec v = \vec r\frac{\dot a}{a}
\quad , \quad \nabla\phi = - \vec r\frac{4\pi G\rho}{3} \quad .
\end{equation}
Notice that these solutions are independent of the pressure. Indeed,
since in a homogeneous Universe $p = p(t)$, there is no pressure gradient,
and the presence of pressure does not influence the evolution of the Universe.
\par
We turn now to the perturbative level. Introducing in equations
(\ref{be1},\ref{be2},\ref{be3}) the quantities
$\rho = \rho_0 + \delta\rho$, $\vec v = \vec v_0 + \delta\vec v$,
$p = p_0 + \delta p$ and $\phi = \phi_0 + \delta\phi$, where each
quantity has been expressed as a sum of the background solution plus
a small perturbation, we can combine the resulting perturbed equation
in order to obtain an unique differential equation for the density
contrast $\frac{\delta\rho}{\rho}$:
\begin{equation}
\label{pe}
\ddot\delta + 2\frac{a}{a}\dot\delta + \biggr\{\frac{v_s^2n^2}{a^2} -
3\frac{\ddot a}{a}\biggl\}\delta = 0 \quad .
\end{equation}
In this expression $v_s^2 = \frac{\partial p}{\partial\rho}$ is the
the sound velocity and $n$ is the wavenumber of the perturbation, which
appears due to the fact that we expanded the spatial dependence of the
perturbed quantities in plane waves.
An important property of the Chaplygin gas is that, in spite of exhibiting
negative pressure, its sound velocity is positive. This does not
happen in general: fluids with negative pressure quite frequently
exhibit instabilities, at least in their hydrodynamical representation,
due to an imaginary sound velocity \cite{jerome,sergio}.
\par
Inserting the background solutions and the expression for the sound velocity in (\ref{pe}),
we determine the behaviour of density perturbations. The solution reads
\begin{equation}
\label{fs}
\delta = t^{-1/6}\biggr[C_1J_\nu(\Sigma^2nt^{7/3})
+ C_2J_{-\nu}(\Sigma^2nt^{7/3})\biggl]
\end{equation}
where $\nu = 5/14$ and $\Sigma^2 = \frac{54}{49}\frac{A\pi G}{a_0^2}$.
In principle $C_1$ and $C_2$ are constants that depend on $n$.
\par
The solution (\ref{fs}) has two asymptotic regime. For
$t \rightarrow 0$, there is a growing mode given by $\delta_+ \propto
t^{2/3}$. This coincides with the evolution of density perturbations in
a dust fluid in General Relativity. On the other hand, for
$t \rightarrow \infty$, the density contrast oscillates with decreasing
amplitude, $\delta_\pm \propto t^{-4/3}\cos(\Sigma^2nt^{7/3} + \phi)$,
where $\phi$ is a phase, approaching a zero value asymptotically. Hence,
for large values of time, the solution approaches the result for
density perturbation in General Relativity with a cosmological constant,
$\delta_{cc} = 0$.
The transition from a "dust" phase to a "cosmological constant"
phase is smooth and the moment it happens is essentially dictated
by the value of the constante $A$ in (\ref{acao}).
This behaviour inverts what happens in the traditional case of density
perturbations in newtonian gravity with positive pressure\cite{weinberg}.
In this case, taking the
equation of state $p \propto \rho^\gamma$, $\gamma > 4/3$, the density
contrast initially oscillates, and after it grows with $\delta_+
\propto t^{2/3}$.
\par
These features have some interesting consequences. In fact, let us
suppose that at the moment of decoupling of radiation and matter, $t_i \sim 10^{11}s$,
the amplitude of density perturbations is fixed by the amplitude
of the anisotropy of cosmic microwave background radiation, i.e.,
$\delta_i \sim 10^{-5}$. Later, deep in the material phase, the
newtonian approximation is reasonable. Supposing the Universe
from that moment on is dominated by a Chaplygin gas fluid (in more
sophisticated words, a gas of p-branes), than density perturbations
grow initially, decreasing later. Observations indicate that the large structure are such today that
$\delta_f < 1$, where the subscript indicates an instant near our now. Supposing the first asymptotic regime for
the density contrast found before is valid,
than 
\begin{equation}
\frac{\delta_f}{\delta_i} = \biggr(\frac{t_f}{t_i}\biggl)^{2/3} \quad .
\end{equation}
The condition $\delta_f < 1$, leads to $t_f \sim 10^{17}s$, that
is $t_f \sim t_0$, where $t_0$ is the age of the Universe.
Hence the approximation that the pressure is negligible must
be valid until very recently in order to explain observations.
This fact limits the value of $A$ as $A < \rho_0$,
where $\rho_0$ is the density of the Universe today.
Only very recently pressure should
become important, and the perturbations have stopped growing.
This fact may have some consequences for the anisotropy of cosmic
microwave background, at least for large multipoles values, that
is, small angles.
\par
In this note, we have outlined the behaviour of density of perturbations
in a specific type of fluid with negative pressure, the Chaplygin gas,
which can be viewed also as a gas of p-branes.
This gas may interpolate a regime of a dust dominated Universe to a
vacuum energy dominated Universe, given an explanation for the
possible accelerated phase today. We performed a newtonian analysis since
the Chaplygin gas obeys the equations of newtonian hydrodynamic.
The main conclusion is that it admits an initial phase of growing
perturbations, with the the same rate as in dust case of the
cosmological standard model, from which it follows decreasing oscillations,
which asymptotically go to zero. Hence, even if the newtonian analysis
is limited for the study of the Universe in large scales, it this
case the newtonian solution interpolates conveniently an initial dust phase ($\delta_d \propto t^{2/3}$)
with a cosmological constant phase ($\delta_{cc} = 0$). Even if the
newtonian background model is very different from the relativistic one,
since it gives a dust-like solution for any value of time, the perturbative
newtonian analysis is consistent with the different phases predicted
by expression (\ref{pressao}), which comes from relativistic considerations.
Moreover, we notice that the Chaplygin gas, in spite of presenting
negative pressure, is stable at small scale, in opposition to what
happens in general with perfect fluids with negative pressure.
This is due to the fact that the sound velocity in the Chaplygin gas
is positive.
\newline
\vspace{0.5cm}
{\bf Acknowledgements:} We thank CNPq (Brazil) for partial financial support.

\end{document}